\documentstyle[12pt,epsf,pennames,epsfig]{article}
\begin{document}


\pagestyle{empty}

\renewcommand{\thefootnote}{\fnsymbol{footnote}}


\begin{flushright}
{\small
SLAC--PUB--7516\\
May 1997\\}
\end{flushright}

\begin{center}
{\large\bf HADRON FRAGMENTATION FUNCTIONS AND } \\[1mm]
{\large\bf LEADING PARTICLE EFFECTS IN HADRONIC Z$^{\bf{0}}$ DECAYS:}\\[1mm]
{\large\bf NEW RESULTS FROM SLD}\footnote{Work supported by
Department of Energy contract  DE--AC03--76SF00515.}

\vspace{1.cm}

{\bf Jochen Schwiening}

Stanford Linear Accelerator Center

Stanford University, Stanford, CA 94309

\vspace{3.mm}
{\it Representing}
\vspace{.2cm}

{\bf The SLD Collaboration}

\end{center}

\vspace{6.cm}   

\begin{center}
{\bf\large   
Abstract }
\end{center}

We have measured the differential cross sections for the production of 
\Pgppm , \PKpm , \PKz , \PKst , \Pgf , \Pp\ and \PgL\ in
hadronic $Z^0$ decays and in subsets of flavor-tagged 
\mbox{$Z^0 \rightarrow$ light} quark ($u \bar{u}$, $d \bar{d}$, or $s \bar{s}$) , 
\mbox{$Z^0 \rightarrow c \bar{c}$} and 
\mbox{$Z^0 \rightarrow b \bar{b}$} events.
Charged hadrons were identified with the SLD Cherenkov ring
imaging detector.  
The vertex detector was employed to select
flavor enriched samples and the polarized electron beam from SLC was 
used to tag quark and anti-quark jets.
We observe a flavor dependence in the hadron fragmentation functions.
We present evidence for leading particle production in hadronic decays of the
$Z^{0}$ boson to light-flavor jets and a direct measurement of the 
strangeness suppression factor $\gamma_s$.

\vfill
\vspace*{1cm}

\begin{center} 
{\it Invited talk presented at the
32nd Rencontres de Moriond:} \\
{\it QCD and High-Energy Hadronic Interactions } \\
{\it Les Arcs, France }\\
{\it 22-29 Mar 1997}\\
\end{center}

\vfill\eject

%
%
\baselineskip=12pt

\vspace*{3.cm}   

\begin{center}
{\large\bf HADRON FRAGMENTATION FUNCTIONS AND } \\[1mm]
{\large\bf LEADING PARTICLE EFFECTS IN HADRONIC Z$^{\bf{0}}$ DECAYS:}\\[1mm]
{\large\bf NEW RESULTS FROM SLD}

\vspace{1.cm}

{\bf Jochen Schwiening}

Stanford Linear Accelerator Center

Stanford University, Stanford, CA 94309

\vspace{3.mm}
{\it Representing}
\vspace{.2cm}

{\bf The SLD Collaboration}

\end{center}

\vspace{8.cm}   

\begin{abstract}
We have measured the differential cross sections for the production of
\Pgppm , \PKpm , \PKz , \PKst , \Pgf , \Pp\ and \PgL\ in
hadronic $Z^0$ decays and in subsets of flavor-tagged
\mbox{$Z^0 \rightarrow$ light} quark ($u \bar{u}$, $d \bar{d}$, or $s \bar{s}$) ,
\mbox{$Z^0 \rightarrow c \bar{c}$} and
\mbox{$Z^0 \rightarrow b \bar{b}$} events.
Charged hadrons were identified with the SLD Cherenkov ring
imaging detector.
The vertex detector was employed to select
flavor enriched samples and the polarized electron beam from SLC was
used to tag quark and anti-quark jets.
We observe a flavor dependence in the hadron fragmentation functions.
We present evidence for leading particle production in hadronic decays of the
$Z^{0}$ boson to light-flavor jets and a direct measurement of the
strangeness suppression factor $\gamma_s$.
\end{abstract}

\clearpage

\baselineskip=18pt    

\noindent {\large \bf 1.   Introduction}\\[-6mm]

The production of hadrons in the decay of the \PZz\ gauge boson
involves the fragmentation stage, that is, the transition of colored partons
into colorless hadrons.
No theoretical description exists yet for this process.
Instead, a variety of phenomenological models has been developed.
At the \PZz\ energies the two most successful models are the string
fragmentation
model, incorporated in the {\sc Jetset} program~\cite{jetset}, and the cluster
fragmentation scheme that is part of the {\sc Herwig} program~\cite{herwig}.
The study of the production of identified hadrons has been, and
is, an important tool~\cite{boehrer} in tests of the fragmentation models
because these particles can often be identified with high purity and
sufficiently large statistics over a wide momentum range.

In this paper we discuss a measurement of the differential cross sections
for the production of
\Pgppm , \PKpm , \PKz , \PKst , \Pgf , \Pp\ and \PgL\ in
hadronic $Z^0$ decays and the first study of leading particle production in
light flavor jets in $e^{+}e^{-}$ annihilation.
The analysis used 150,000 hadronic $Z^{0}$ decay events produced by the
SLAC Linear
Collider (SLC) and recorded in the SLC Large Detector (SLD) from 1993
to 1995.
%
%
%

A description of the SLD detector, trigger, track and hadronic event
selection, and Monte Carlo simulation is given in Ref.~\cite{impact}.
Cuts were applied in order to select events well-contained within the
detector acceptance, resulting in a sample of approximately 90,000 events.
Sub-sets of flavor-tagged
$e^{+}e^{-}\rightarrow Z^{0}\rightarrow q\bar{q}$, where $q = u, d,$ or $s$,
and
$e^{+}e^{-}\rightarrow Z^{0}\rightarrow Q\bar{Q}$, where $Q = c$ or $b$
were selected by using information from the Vertex
Detector (VXD)~\cite{vxd} and
applying an impact parameter technique~\cite{thepaper}.\\[-4mm]

\noindent {\large \bf  2.   Particle Identification}\\[-6mm]

The identification of $\pi^{\pm}$, \PKpm , p, and $\bar{\mathrm p}$ was achieved
by reconstructing emission angles of individual Cherenkov photons radiated by
charged particles passing through liquid and gas radiator systems of the SLD
Cherenkov Ring Imaging Detector (CRID)~\cite{crid}.
In each bin of the scaled momentum $x_{p} =  2p/\sqrt{s}$ of the
hadron, where $p$ is its magnitude of momentum and $\sqrt{s}$ is the
$e^{+}e^{-}$ center-of-mass energy,
identified $\pi$, $K$, and p were
counted, and these counts were unfolded using the inverse of the identification
efficiency matrix ${\bf E}$~\cite{partid,pavel}, and corrected for track
reconstruction efficiency.
The elements $E_{ij}$, denoting the momentum-dependent
probability to identify a true $i$-type particle as a $j$-type particle, were
measured from the data for $i=\pi ,{\mathrm p}$ and $j=\pi ,K,{\mathrm p}$ using tracks
from selected \PKzS , $\tau$ and $\Lambda$ decays.  A detailed Monte Carlo
simulation was used to derive the remaining elements in terms of these measured
ones.

Candidate $\PKzS\rightarrow\Pgpp\Pgpm$, $\Lambda\rightarrow{\mathrm p}\pi^{-}$ and
$\bar{\Lambda}\rightarrow\bar{\mathrm p}\pi^{+}$ decays were selected by
considering all pairs of oppositely charged tracks that were inconsistent with
originating at the interaction point and passed a set of cuts~\cite{baird} on
vertex quality and flight distance.
Backgrounds from mis-identified $\Lambda$ and \PKzS\ decays and
photon conversions were suppressed by using kinematic cuts.

Candidate $K^{*0}\rightarrow K^+\pi^-$ /
$\overline{K}^{*0}\rightarrow K^-\pi^+$ decays
were selected by considering all pairs of oppositely-charged tracks in
which exactely one track was identified in the CRID as
a charged kaon,
and the tracks were consistent with intersecting at the interaction
point~\cite{dima}.
\newpage
Candidate $\Pgf\rightarrow\PKp\PKm$ decays
were selected by considering all pairs of oppositely-charged tracks in
which both tracks were identified in the CRID as charged kaons,
and the tracks were consistent with intersecting at the interaction
point~\cite{dima}.

In each $x_{p}$ bin, the number of observed
\PKz / \PaKz\ ,  $\Lambda /\bar{\Lambda}$,  $K^{*0} / \overline{K}^{*0}$
and \Pgf\
was determined from a fit to the appropriate invariant mass distribution.
Finally, the signals were corrected for reconstruction efficiencies.
\\[-4mm]

\begin{figure}[thb]
\hspace*{-10mm}
  \epsfxsize=1.1\textwidth
  \epsfig{figure=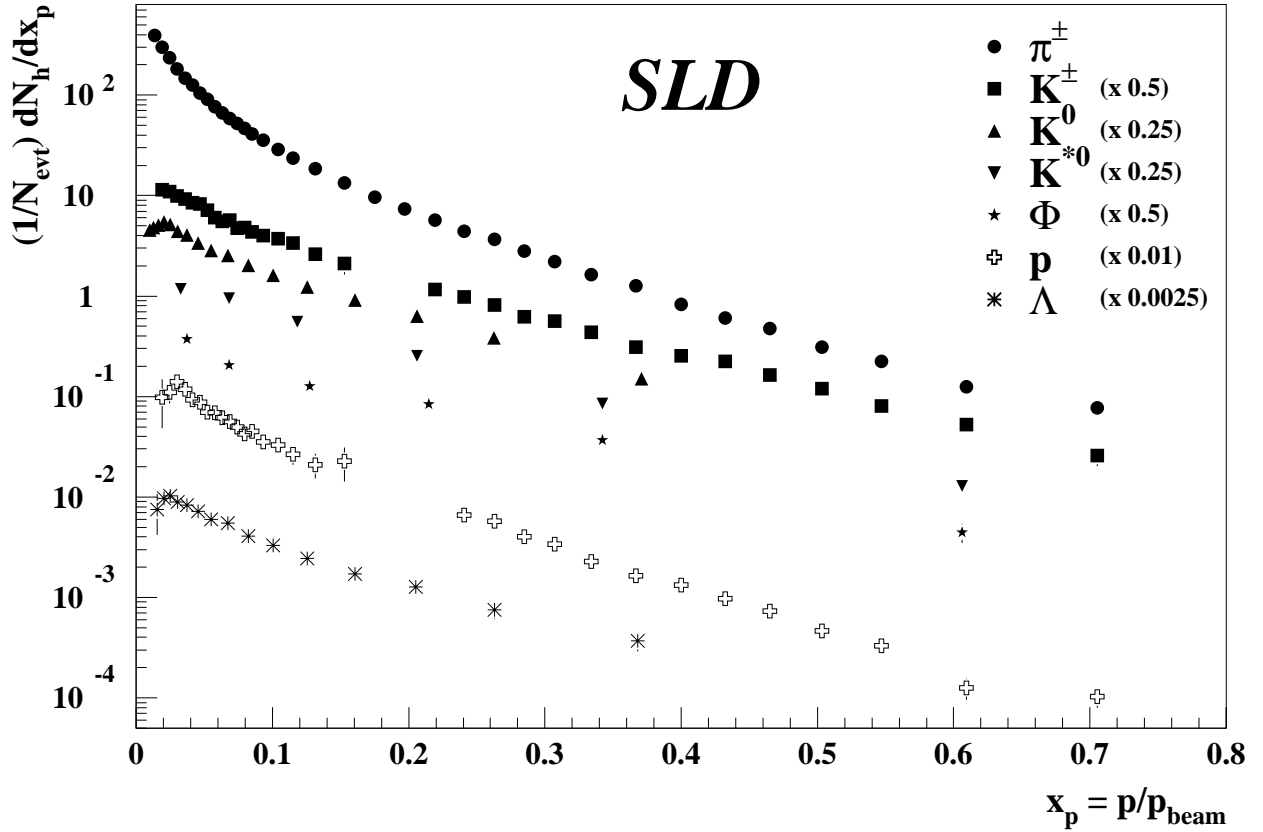}
  \caption{\it \label{fig:xsect}
  \baselineskip=12pt
  Differential cross sections in inclusive \PZz\ decays
  as a function of scaled momentum.
  The errors shown are combined statistical and bin-by-bin systematic uncertainties.}
  \baselineskip=18pt
\end{figure}

\noindent {\large \bf  3.   Hadron Fragmentation Functions}\\[-6mm]

\begin{figure}[thb]
\hspace*{-10mm}
  \epsfxsize=1.1\textwidth
  \epsfig{figure=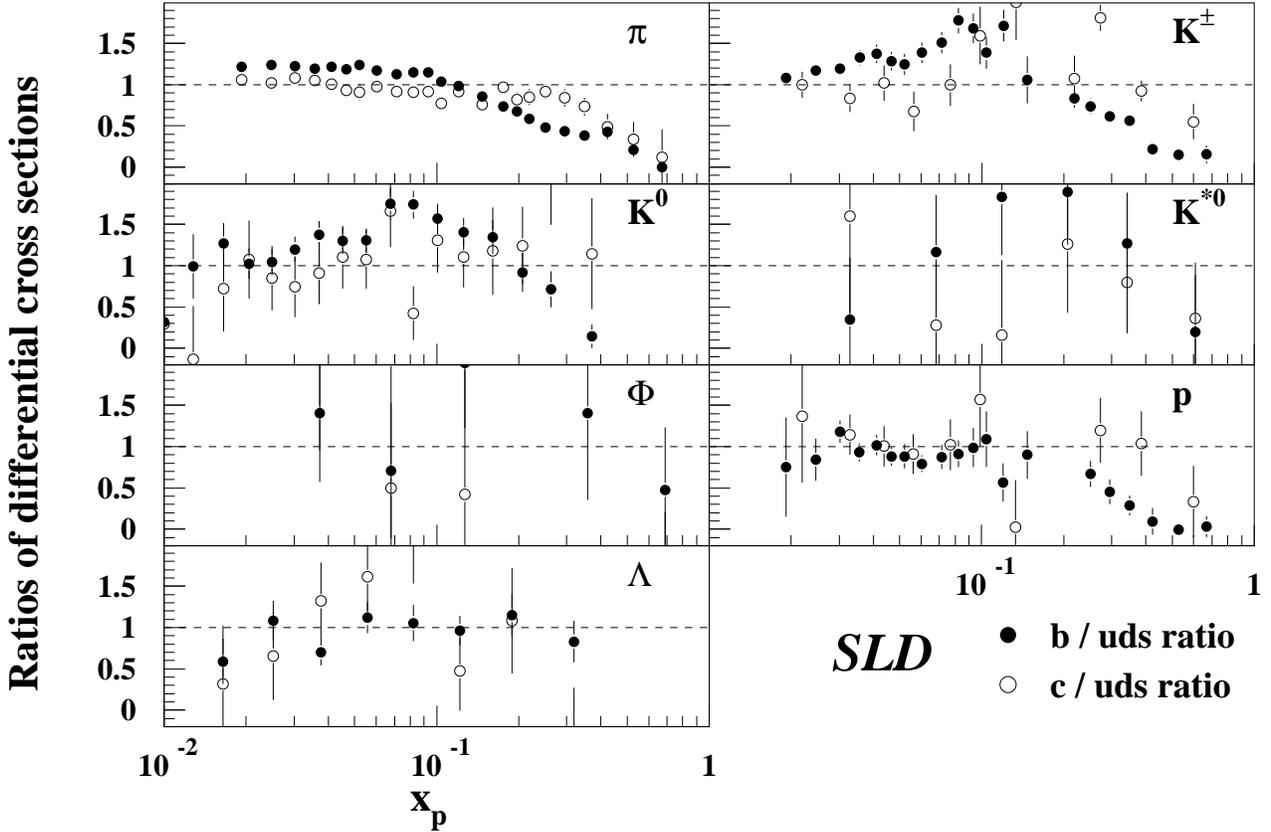}
  \caption{\it \label{fig:xsect_flav}
  \baselineskip=12pt
  Ratios of differential cross sections in flavor tagged \PZz\ decays
  as a function of scaled momentum.
  The errors shown are combined statistical and bin-by-bin systematic uncertainties.}
  \baselineskip=18pt
\end{figure}

The differential cross sections for
\Pgppm , \PKpm , \PKz , \PKst , \Pgf , \Pp\ and \PgL\ in
inclusive hadronic $Z^0$ decays are shown in Fig.~\ref{fig:xsect}
as a function of $x_{p}$.
There are no $K^{\pm}$
or p/$\bar{\rm p}$ points in the range $0.12 < x_p < 0.20$ due to the lack of
CRID particle separation in this region.
We also determined the hadron fragmentation functions in sub-samples of
flavor-tagged light quark ($u \bar{u}$, $d \bar{d}$, or $s \bar{s}$),
$c \bar{c}$ and $b \bar{b}$ events.
The ratios of the differential cross sections in $c \bar{c}$ and $b \bar{b}$
events to light quark events are shown in Fig.~\ref{fig:xsect_flav}.
A clear flavor dependence can be seen.
In particular, the production of the mesons is enhanced in $b$-jets for
momenta below a few GeV/c.
At higher momentum, the light hadrons are predominantly produced in
$u \bar{u}$, $d \bar{d}$, or $s \bar{s}$ jets.
Even though the errors are large, it can be seen that the value of $x_p$
at which the production in light flavor jets becomes dominant is larger
for $c$-quark jets than for bottom events, as expected from
the heavy hadron decay properties.
\\[-4mm]

\pagebreak
\noindent {\large \bf  4.   Leading Particle Effects}\\[-6mm]

We define a particle to be leading if
it carries a primary quark or antiquark, namely the $q$ or $\bar{q}$ in
$e^{+}e^{-}\rightarrow Z^{0}\rightarrow q\bar{q}$.
We separated jets initiated by primary quarks from those initiated by primary
antiquarks by utilizing the electroweak forward-backward production asymmetry
in the polar angle, enhanced by the high SLC electron beam polarization.
We considered all events to consist of one jet in each of the two
hemispheres separated by the plane perpendicular to the thrust axis.
Defining
the forward direction to be along the electron beam, the quark jet was defined
to comprise the set of tracks in the forward (backward) hemisphere for events
recorded with left-(right-) handed electron beam.  The opposite jet in each
event was defined to be the antiquark jet.
For details on the tagging procedure, see Ref.~\cite{thepaper}.
In our sample of approximately 41,000 light quark
events we measured the differential production rates per light quark jet
(as opposed to light anti-quark jet)
\begin{eqnarray*}
R^{q}_{h} &=& {1\over{2N_{evts}}}{d\over{dx_{p}}}\left[ N(q\rightarrow
h)+N(\bar{q}\rightarrow\bar{h})\right],\\
R^{q}_{\bar{h}} &=& {1\over{2N_{evts}}}{d\over{dx_{p}}}\left[
N(q\rightarrow\bar{h})+N(\bar{q}\rightarrow h)\right],
\end{eqnarray*}
where: $q$ and $\bar{q}$ represent light-flavor quark and anti-quark jets
respectively; $N_{evts}$ is the total number of events in the sample; $h$
represents any of the identified hadrons $\pi^{-}$, $K^{-}$,
$\overline{K}^{*0}$, p, and $\Lambda$, and $\bar{h}$ indicates the
corresponding anti-particle;
Then, for example, $N(q\rightarrow h)$ is
the number of hadrons of type $h$ in light quark jets.
In every $x_{p}$ bin, $R^{q}_{h}$ and $R^{q}_{\bar{h}}$ were
corrected for the contribution from residual heavy-flavor events,
estimated from our Monte Carlo simulation.
Finally, the corrected $R^{q}_{h}$
and $R^{q}_{\bar{h}}$ were unfolded for the purity of the quark jet tag.

\begin{figure}[thb]
\hspace*{-10mm}
  \epsfxsize=1.1\textwidth
  \epsfig{figure=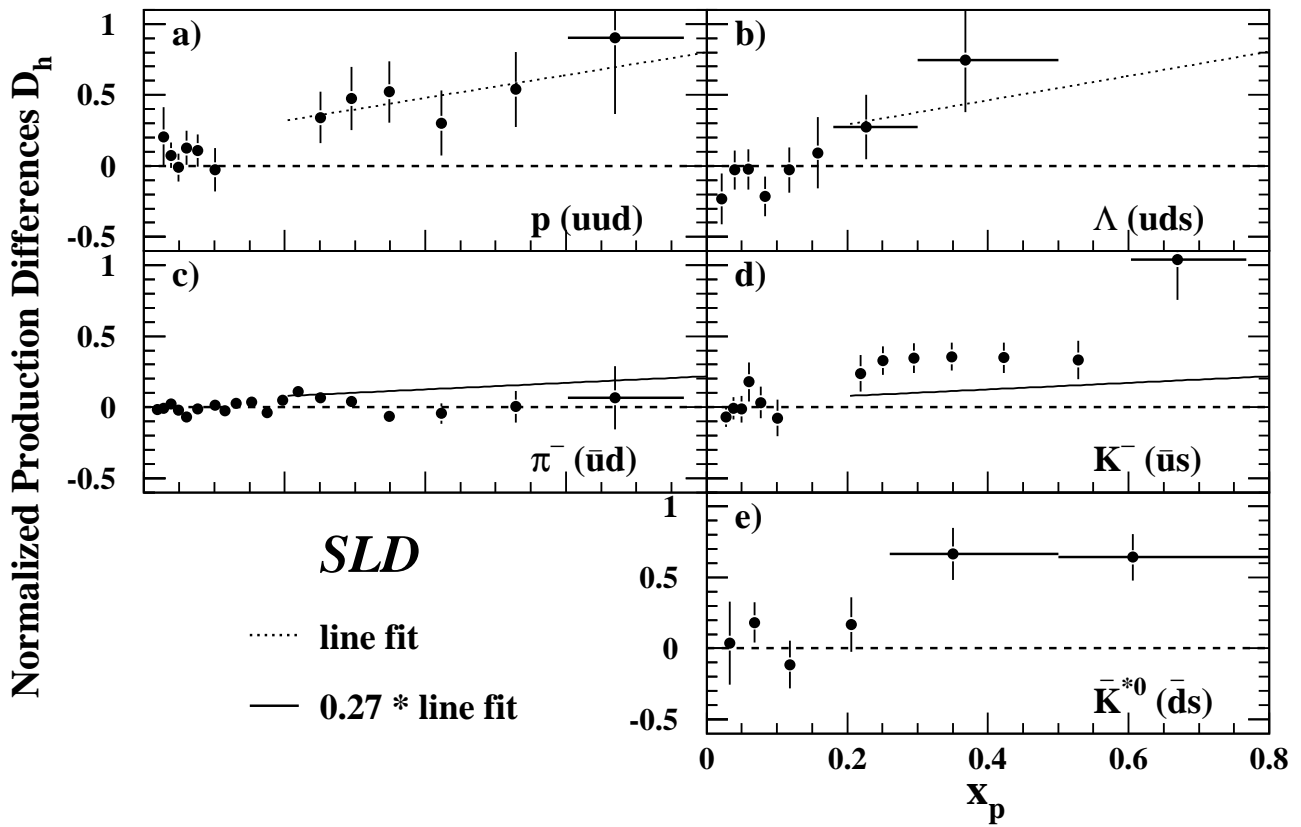}
  \caption{\it \label{fig:d_h}
  \baselineskip=12pt
   Normalized production differences (dots) as a function of scaled momentum.
   The vertical error bars shown are combined statistical and bin-by-bin systematic
   uncertainties.
   The horizontal error bars on selected points
   indicate their bin widths.  The dotted lines represent a linear fit to the
   $D_{p}$ and $D_{\Lambda}$ points for $x_p > 0.2$, and the solid lines
   are this fit scaled by the factor 0.27 discussed in the text.}
  \baselineskip=18pt
\end{figure}

We define the difference between each particle and anti-particle production
rate, normalized by the sum: $$D_{h} =  {R^{q}_{h} - R^{q}_{\overline{h}}\over
R^{q}_{h} + R^{q}_{\overline{h}}},$$ for which the common systematic
uncertainties cancel.  As shown in Fig.~\ref{fig:d_h}, for each hadron $h$,
$D_{h}$ is consistent with zero for $x_p < 0.1$.  $D_{\pi^{-}}$ is also
consistent with zero for $x_p > 0.1$, but for the other hadrons $D_{h} > 0$ for
$x_p$ \raisebox{-.7ex}{$\stackrel{\textstyle >}{\sim}$} 0.2.  The JETSET
7.4~\cite{jetset} and HERWIG 5.8~\cite{herwig} fragmentation models were found
to reproduce these features qualitatively.

Since baryons contain no constituent anti-quarks, we interpret the positive
$D_{p}$ and $D_{\Lambda}$ as evidence for leading baryon production in
light-flavor jets.
If pions and kaons exhibited similar leading effects, then,
due to the larger asymmetry parameter for u-type than d-type quarks,
one would expect $D_{\pi^-} \approx D_{K^-}  \approx  0.27 D_{baryon}$, and
$D_{\overline{K}^{*0}} = 0$, assuming Standard Model quark couplings to the
$Z^0$.
For purposes of illustration, the result of a linear fit to the $D_{p}$
and $D_{\Lambda}$ points above $x_p=0.2$ was scaled by 0.27 and is shown in
Figs.~\ref{fig:d_h}(c) and \ref{fig:d_h}(d).
The observed $D_{\pi^-}$ are below this line, and are
consistent with zero at all $x_p$, suggesting that either there is little
production of leading pions, or there is substantial background from
non-leading pions or pions from decays of resonances such as the $\rho$ and
$K^*$.
For $x_p > 0.2$, we observe $D_{K^-} > 0.27 D_{baryon}$ and
$D_{\overline{K}^{*0}} > 0$.  This indicates both substantial production of
leading $K$ and $K^*$ mesons at high momentum, and a depletion of leading kaon
production in $u\bar{u}$ and $d\bar{d}$ events relative to $s\bar{s}$ events.

Assuming these high-momentum kaons to be directly produced in the fragmentation
process, this amounts to a direct observation of a suppression of $s\bar{s}$
production from the vacuum with respect to $u\bar{u}$ or $d\bar{d}$ production.
In the case of $K^{*0}$ mesons it has been suggested~\cite{lafferty} that this
effect can be used to measure the ``strangeness suppression parameter''
$\gamma_s$, that is an important component of models of hadronization, see e.g.
Ref.~\cite{jetset}.  Assuming {\it all} $K^{*0}$ and $\overline{K}^{*0}$ in the
range $x_{p}>0.5$ to be leading, we calculate $\gamma_s = 0.26\pm 0.12$,
where the error is predominantly statistical,
consistent with values~\cite{boehrer} derived from inclusive measurements of the
relative production rates of strange and non-strange, pseudoscalar and vector
mesons. \\[-4mm]

\noindent {\large \bf       Acknowledgments}\\[-6mm]

It is a pleasure to thank the organizers of the conference for
their efforts in arranging an enjoyable meeting.
We thank the personnel of the SLAC accelerator department
and the technical staffs of our collaborating institutions for their
outstanding efforts.
I would like to express my thanks to my SLD colleagues K.~Baird, P.~Burrows,
M.~Dima, M.~Kalelkar, D.~Muller and T.~Pavel
for their help in the preparation of this talk and manuscript.
Finally, I am grateful to the Alex\-ander-\-von-\-Hum\-boldt Stiftung
for their financial support.
\\[-4mm]

\end{document}